\documentclass[conference]{IEEEtran}
\ifCLASSINFOpdf
\else
\fi
\usepackage{amsfonts}
\usepackage{amssymb}
\usepackage{amsmath,bm}
\usepackage{graphicx}
\usepackage{algorithm}
\usepackage{algorithmic}
\usepackage{multirow}
\usepackage{amsmath}
\usepackage{xcolor}
\usepackage{subfigure}
\usepackage{url}

\begin{document}
%
\title{An Empirical Analysis of Task Allocation in Scrum-based Agile Programming}



%
\author{
\IEEEauthorblockN{Jun Lin\IEEEauthorrefmark{1}\IEEEauthorrefmark{2},
Han Yu\IEEEauthorrefmark{1} and
Zhiqi Shen\IEEEauthorrefmark{1}}

\IEEEauthorblockA{\IEEEauthorrefmark{1}School of Computer Engineering, Nanyang Technological University, Singapore}
\IEEEauthorblockA{\IEEEauthorrefmark{2}School of Software, Beihang University, Beijing, China}

\IEEEauthorblockN{\{jlin7, han.yu, zqshen\}@ntu.edu.sg, linjun@buaa.edu.cn}
}


\maketitle

\begin{abstract}
Agile Software Development (ASD) methodology has become widely used in the industry. Understanding the challenges facing software engineering students is important to designing effective training methods to equip students with proper skills required for effectively using the ASD techniques. Existing empirical research mostly focused on eXtreme Programming (XP)-based ASD methodologies. There is a lack of empirical studies about Scrum-based ASD programming which has become the most popular agile methodology among industry practitioners. In this paper, we present empirical findings regarding the aspects of task allocation decision-making, collaboration, and team morale related to the Scrum ASD process which have not yet been well studied by existing research. We draw our findings from a 12 week long course work project in 2014 involving 125 undergraduate software engineering students from a renowned university working in 21 Scrum teams. Instead of the traditional survey/interview based methods, which suffer from limitations in scale and level of details, we obtain fine grained data through logging students' activities in our online agile project management (APM) platform - \textit{HASE}. During this study, the platform logged over 10,000 ASD activities. Deviating from existing preconceptions, our results suggest negative correlations between collaboration and team performance as well as team morale.
\end{abstract}


%
\IEEEpeerreviewmaketitle

\section{Introduction}
Over the past decade, many companies have fundamentally changed the way they tackle challenges in the software engineering. In place of the traditional plan-driven approaches, the agile software development (ASD) approaches are becoming increasingly widely adopted \cite{Dyba-et-al:2008,Agile:2013}. ASD embodies a new way of thinking and working. It alters the philosophy and practice governing how companies collaborate with their customers, how companies coordinate development, and how engineers develop software. ASD values a set of core principles such as iterative shipments of working software increments, early attention into software quality by all developers, close customer collaboration for fast feedback, and a focus on collaboration within a software development team \cite{Fowler-et-al:2001}. Many different agile methodologies (e.g., Scrum, eXtreme Programming (XP)) have been proposed to help developers implement these agile principles in their daily work.

In recent years, there has been a growing interest to understand how well software developers adapt to ASD \cite{Dyba-et-al:2007,Hannay-et-al:2010,Salleh-et-al:2011,Lin:2013}. Nevertheless, to date, research in this field remains limited and more empirical studies are constantly sought after by researchers. According to the 8th annual State of Agile Survey in 2013 \cite{Agile:2013}, Scrum and its variants are the most popular ASD methodologies (adopted by 73\% of practitioners). Scrum is an iterative and incremental ASD methodology. It consists of six phases: 1) \textit{conceptualization} defines the high level deliverables and project roadmap; 2) \textit{release planning} assigns deliverables into different releases; 3) \textit{sprint planning} breaks down selected deliverables into technical tasks; 4) in each \textit{sprint} (e.g., a 7 day period), software development tasks are to be completed by ASD team members; 5) in \textit{sprint review/retrospective}, team members demonstrate the product increments and reflect on experience gained from the last sprint; and 6) during \textit{release} the working software is delivered to the customers \cite{Chow-Cao:2008}. However, most of the existing empirical research focused on studying the XP methodology instead of the Scrum methodology \cite{Dingsoyr-et-al:2012}. This dichotomy between research effort and industry practice results in ASD training providers lacking important insight into the Scrum ASD process for effectively training prospective software engineers.

In this paper, we attempt to bridge this research gap with a 12 week field study involving 125 second year undergraduate software engineering students from March to June 2014. The students were new to the Scrum methodology and self-organized into 21 Scrum teams of 5 to 7 persons each. This study provides a unique opportunity to study the impact of the ASD methodology used in Scrum teams. Students in this study carried out activities at various stages of the Scrum methodology in our online agile project management (APM) tool - the Human-centred Agile Software Engineering (HASE) platform\footnote{\url{http://www.linjun.net.cn/hase/}}. The activities for each team member supported by HASE mainly occur during the \textit{sprint planning} and \textit{sprint review/retrospective} phases. They include proposing tasks, estimating the priority, difficulty and time required for each task, deciding how to allocate tasks, collaboration information, reviewing the timeliness and quality of completed tasks, and providing feedback about each team member's mood at different points in time during a sprint. During the study, students logged 10,779 ASD activities in the HASE platform.

In our previous round of study in 2013 \cite{Lin-et-al:2014}, this new form of ASD activity data has been shown to produce insights into the ASD process traditional survey/interview based approaches are unable to achieve. The results from this paper offer additional insights into ASD task allocation decision-making, collaboration, and team morale which, to the best of our knowledge, have not been reported by published research study before. Specifically, the results point towards strong positive correlations between a student's technical productivity and the amount of workload allocated to him/her. In addition, contrary to popular preconceptions, collaboration among student team members who are new to Scrum has shown negative correlation with team performance and team morale.

The rest of this paper is organized as follow. Section \ref{st:related-work} outlines related work in studying the ASD process through both survey-based approaches and activity data-based approaches. In Section \ref{st:study-design}, we discuss the design of our study, the metrics we measure, and the characteristics of the student participants. Section \ref{st:results} presents key empirical findings from our study. The limitations of the study in terms of internal, external, and construct validity are discussed in Section \ref{st:limitations}. Section \ref{st:conclusions} concludes the paper and points to potential future research directions.

\section{Related Work} \label{st:related-work}
In \cite{Salleh-et-al:2011}, the authors present the results of a systematic literature review concerning agile pair programming effectiveness.
The paper analyzed compatibility factors, such as the feel-good, personality, and skill level factors, and their effect on pair programming effectiveness as a pedagogical tool in Computer Science and Software Engineering education. Four metrics were used in the analysis: 1) academic performance, 2) technical productivity, 3) program/design quality and 4) learning satisfaction. The general findings are that pair programming is more effective in terms of technical productivity, learning satisfaction and academic performance, while not significantly different in terms of program quality as compared to solo programming. While our study also look into similar metrics, our data are in the form of ASD activity logs. In addition, our study focuses on the Scrum methodology instead of XP.

A number of studies about the Scrum ASD processes start to emerge in recent years. In \cite{Zanniera-et-al:2007}, the authors investigate decisions related to software designs. They employ content analysis and explanation building to extract qualitative and quantitative results from interviews with 25 software designers. The study finds that the structure of the design problem determines the designers choice between rational and naturalistic decision making.

The study in \cite{Drury-et-al:2011} focuses on decision-making by Scrum teams. It examines decisions made across the four stages of the sprint cycle: sprint planning, sprint execution, sprint review and sprint retrospective. The authors employ the technique of focus group study with 43 Agile developers and managers to determine what decisions are made at different points of the sprint cycle. In another publication by the same research group \cite{Drury-et-al:2012}, interviews with an additional 18 professional Agile practitioners from one global consulting organization, one multinational communications company, two multinational software development companies, and one large museum organization are analyze to identify six key obstacles facing decision making in Agile teams.

Nevertheless, the techniques used by existing studies mainly involve interviews and surveys. This limited the scale of the study as well as the level of details of the collected data. As a result, the form of findings from such studies tend to be qualitative in nature. For instance, in \cite{Drury-et-al:2011}, some obstacles facing agile teams during sprint decision-making can ``people are unwilling to commit to a decision", ``lack of ownership" and ``lack of empowerment". There is a lack of quantitative results indicating the extent of each obstacle facing agile team members with different competence levels.

The need to balance the workload among ASD team members while making full use of the more competent of them has been inspired by similar problems facing crowdsourcing systems \cite{Yu-et-al:2012}. The Scrum master, in this case, need to coordinate ASD team members to make efficient task allocation decisions. In \cite{Lin-et-al:2014}, Scrum activity data from an APM platform are analyzed to study students' behavior tendencies when making task allocation decisions. The study uses the same research techniques as reported in this paper. However, the empirical results from this paper involve additional dimensions of data about team collaboration and team morale, and produce new insights into these aspects of the ASD process which are not yet well studied.

\section{Study Design} \label{st:study-design}
In this section, we present our research approach, the metrics that have been measured, and the key characteristics of the student population involved in the study.

\subsection{Research Approach}
Our goal is to investigate the aspects of decision-making, collaboration, and team morale in the Scrum ASD process as practised by student developers who are new to Scrum in the natural settings where these activities occur. Therefore, students in this study perform Scrum ASD activities in our HASE online APM platform. The platform provides five main features to support agile project management which cover the \textit{sprint planning} and \textit{sprint review/retrospective} phases:
\begin{enumerate}
  \item \textit{Registration}: In order to build user profiles, HASE requires registrants to specify their self-assessed competence levels in different areas of expertise such as familiarity with specific programming languages, system design methodologies, and user interface (UI) design tools, etc.
  \item \textit{Team and Role Management}: HASE supports the creation of teams, the selection of product owners and stakeholders into the teams, and the assignment of different roles within a team (e.g., programmers and UI designers).
  \item \textit{Task Management}: Task information including task description, skills required for the task, and the person who proposed each task is displayed for all team members to view. The difficulty value of each task $\tau$, is recorded using an 11-point Likert scale \cite{Likert:1932} (with 0 denoting ``extremely easy" and 10 denoting ``extremely hard"). Each team member can input his/her estimated difficulty value for each task into the HASE platform. The HASE platform then uses the average difficulty value for the task ($D_{\tau}$). The students were asked to take into account the technical challenge as well as the amount of effort required when judging the difficulty of a task. The priority value of each task $\tau$, is also recorded using an 11-point Likert scale (with 0 denoting ``extremely low priority" and 10 denoting ``extremely high priority"). Each team member can input his/her estimated priority value for each task into the HASE platform. The HASE platform then uses the average priority value for the task ($Prio_{\tau}$).
  \item \textit{Sprint Planning}: HASE records the teams' decisions on which tasks are assigned to which team member during each sprint. Once assigned, the status of the task becomes ``Assigned". The assignee $i$ inputs his/her confidence value ($Conf_{\tau}^{i}$) for each task $\tau$ on an 11-point Likert scale (with 0 denoting ``not confident at all" and 10 denoting ``extremely confident"). Each team member also inputs the estimated required time to complete each task (in number of days). The HASE platform uses the average estimated time required to generate the deadline for the task ($T_{\tau}^{est}$). Apart from a primary assignee, multiple students can collaboratively work on a task. The collaborator information for each task is also recorded by HASE.
  \item \textit{Sprint Review/Retrospective}: Once a task is completed, the assignee changes its status in the HASE platform to ``Completed". This action will trigger HASE to record the actual number of days ($T_{\tau}^{act}$) used to complete this task. HASE also provides functions for team members to peer review the quality ($Qual_{\tau}$) of each completed task $\tau$. The quality of a completed task is recorded in the platform using a 11-point Likert scale with 0 representing (``extremely low quality") and 10 representing (``extremely high quality"). The average quality rating for each task is used by HASE as the final quality rating for that task.
  \item \textit{Team Morale Monitoring}: During the sprint planning meeting, team members can report their current mood values into the HASE platform. A person $i$'s mood at the beginning of a sprint $t$ ($m_{i}^{begin}(t)$) is represented on a 5-point Likert scale with 1 representing ``very low" and 5 representing ``very high". During the sprint review/retrospective meeting, each task assignee $i$ can report his/her mood after completing a task at the end of sprint $t$ ($m_{i}^{end}(t)$) using the same 5-point Likert scale.
\end{enumerate}

The input data to the HASE platform required from ASD teams are as a result of students' activities following the Scrum methodology. In this way, users of HASE can behave as if they are using any APM tool without expending additional effort to help with data collection. Thus, the data collection process remains unobtrusive to the participants

A total of 125 second year undergraduate software engineering students from the Beihang University, China were involved in this study. The students need to form into Scrum teams of 5 to 7 persons each to complete a team-based software engineering project over a 12 week period of time. As this is part of the students' course work, the curriculum dictates that the students must decide among themselves how to form into teams. Eventually, the students formed into 21 teams. Each team then proposes a software engineering project for the course instructor to approve. The projects are mediated by the instructor so that they are of comparable scale and complexity across all teams. Some examples of the proposed software projects in this study are ``An android system for interest-based music recommendation", ``A mobile health information app for the elderly", ``A mobile app for monitoring user mobility pattern", etc. The teams then adopt the Scrum process to develop their projects. Each sprint lasts a week. During the sprint planning meeting, team members propose the tasks that need to be completed over this sprint. A total of 893 tasks have been proposed by all teams during this study. Students perform all Scrum ASD activities using the HASE platform.

\subsection{Metrics}

\begin{figure*}[t]
    \centering
    \subfigure[The distribution of students' capabilities in the study]{
    \includegraphics[trim = 40mm 85mm 45mm 90mm, clip, width = 2.2in]{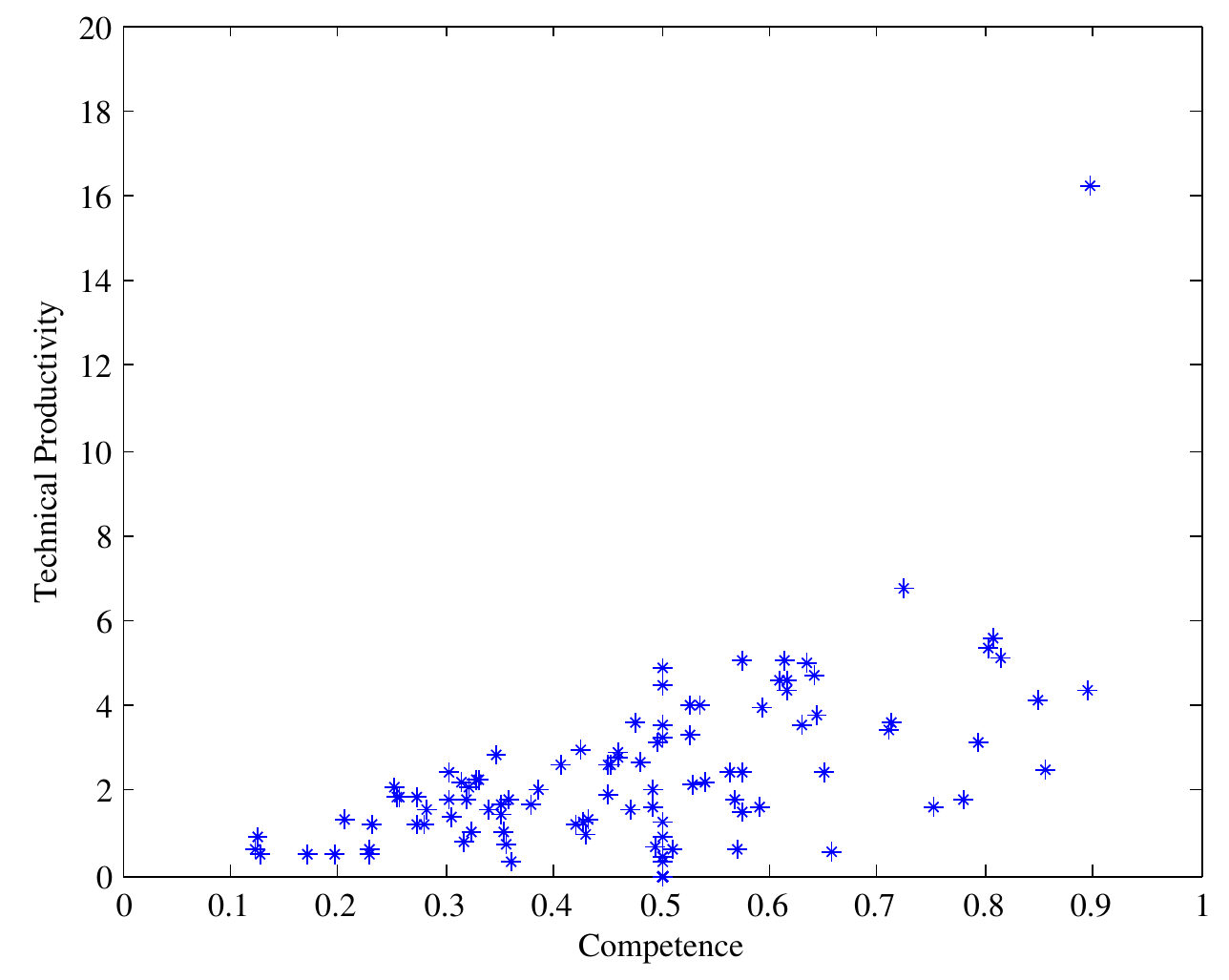}\label{fig:StudentsCompCap}}
    \subfigure[The distribution of teams' capabilities in the study]{
    \includegraphics[trim = 40mm 85mm 45mm 90mm, clip, width = 2.2in]{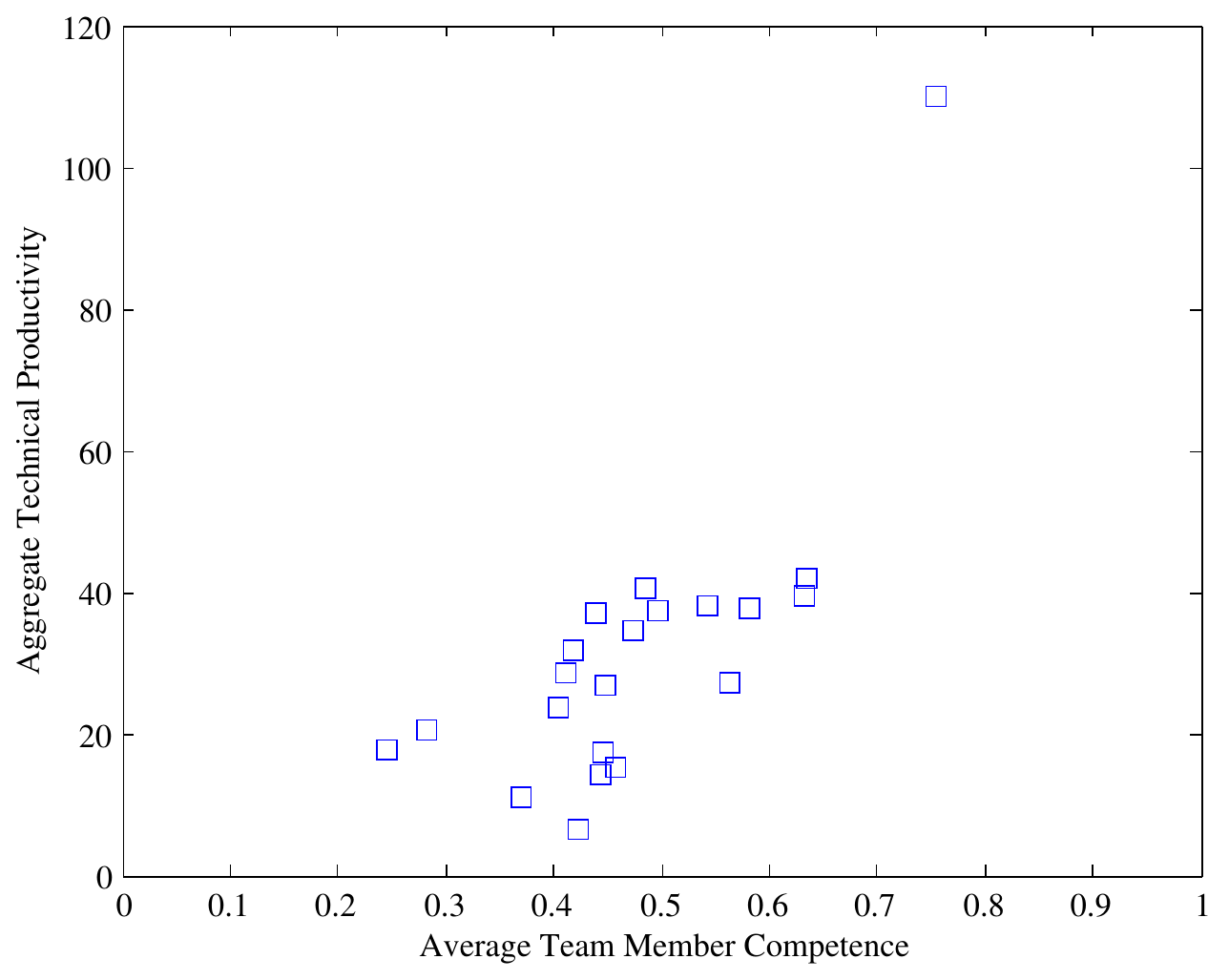}\label{fig:TeamsCompCap}}
    \subfigure[The distribution of students' final scores]{
    \includegraphics[trim = 40mm 85mm 45mm 90mm, clip, width = 2.2in]{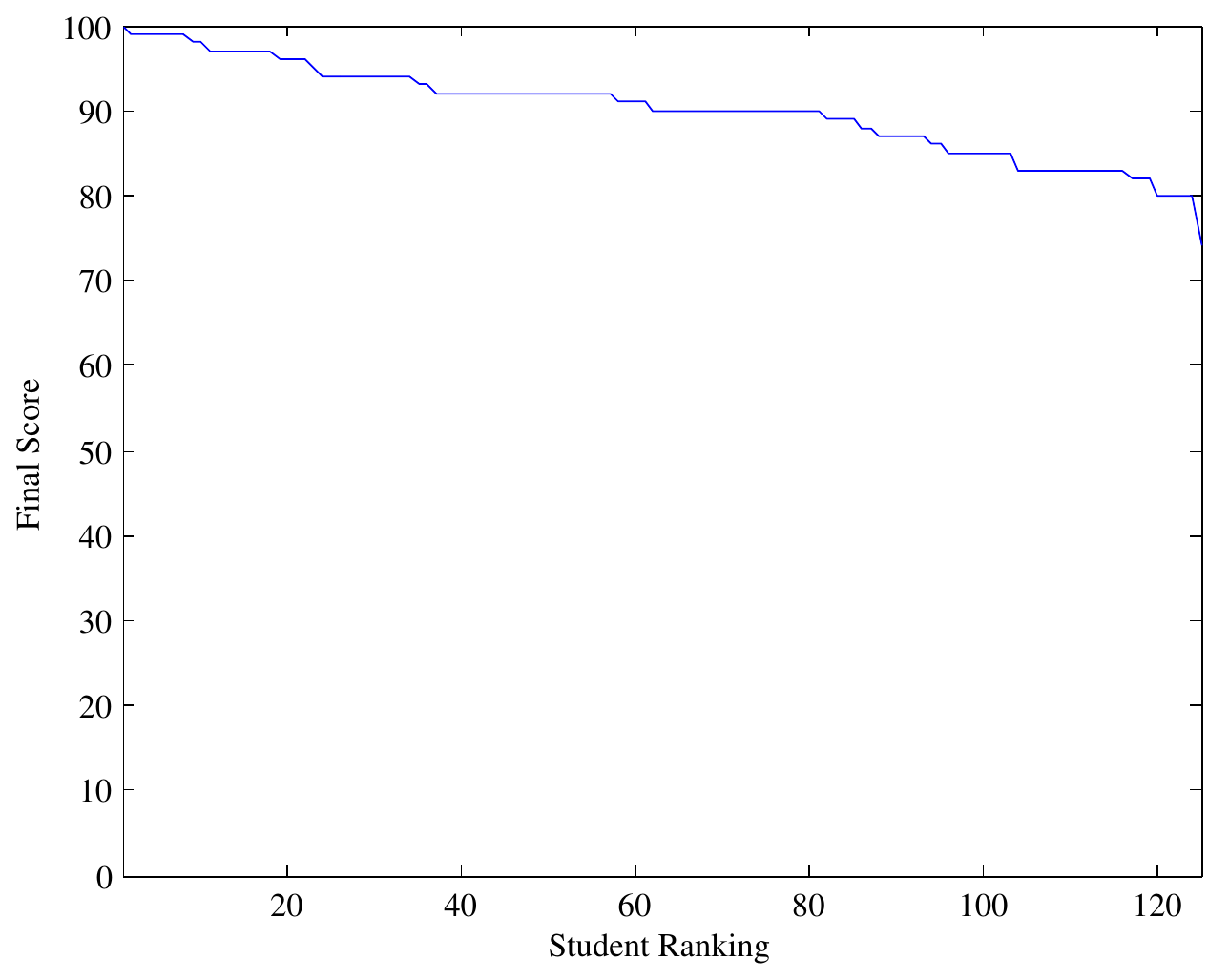}\label{fig:FinalScores}}
    \subfigure[The distribution of students' competence]{
    \includegraphics[trim = 40mm 85mm 45mm 90mm, clip, width = 2.2in]{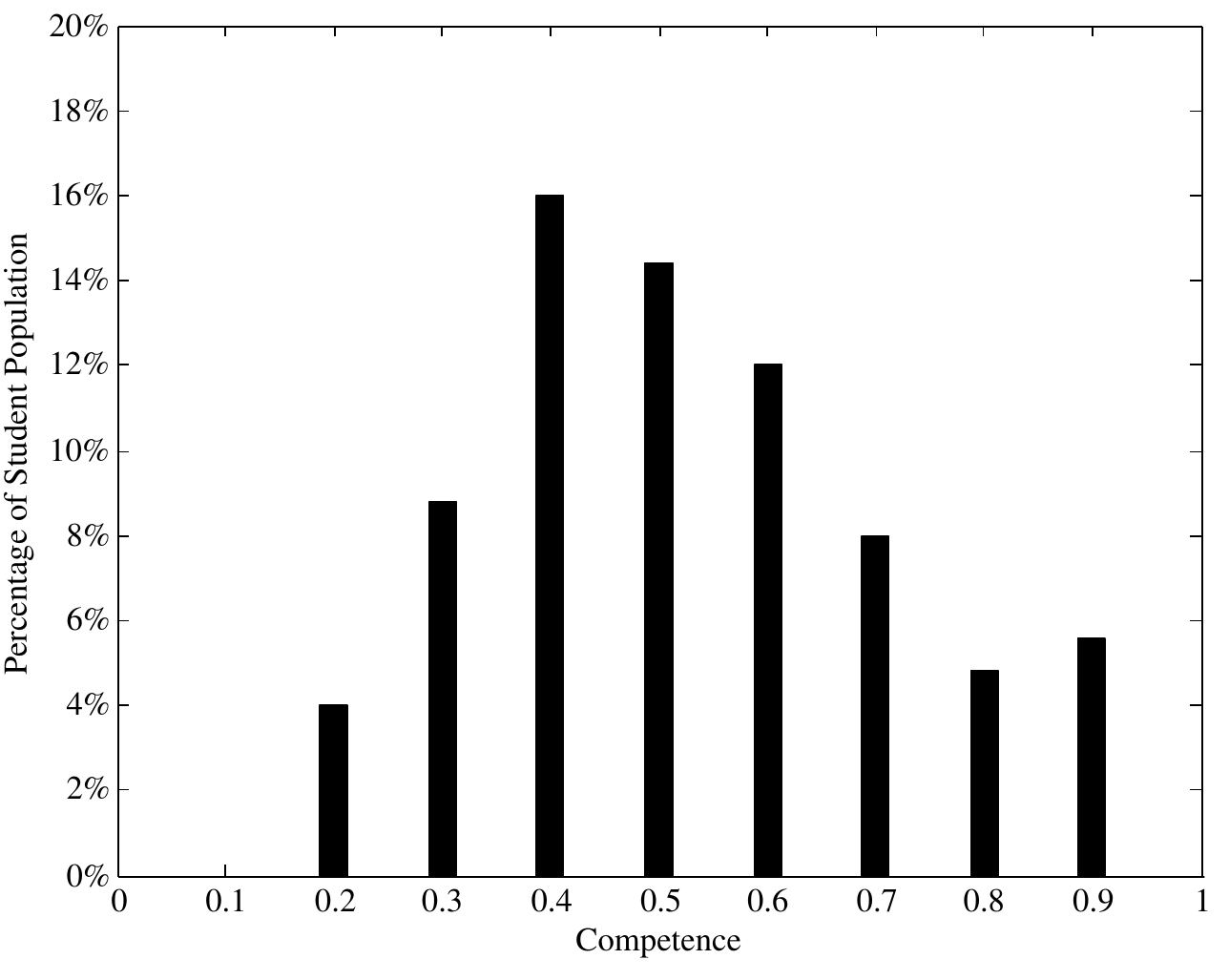}\label{fig:CompDistribution}}
    \subfigure[The distribution of task difficulty values]{
    \includegraphics[trim = 40mm 85mm 45mm 90mm, clip, width = 2.2in]{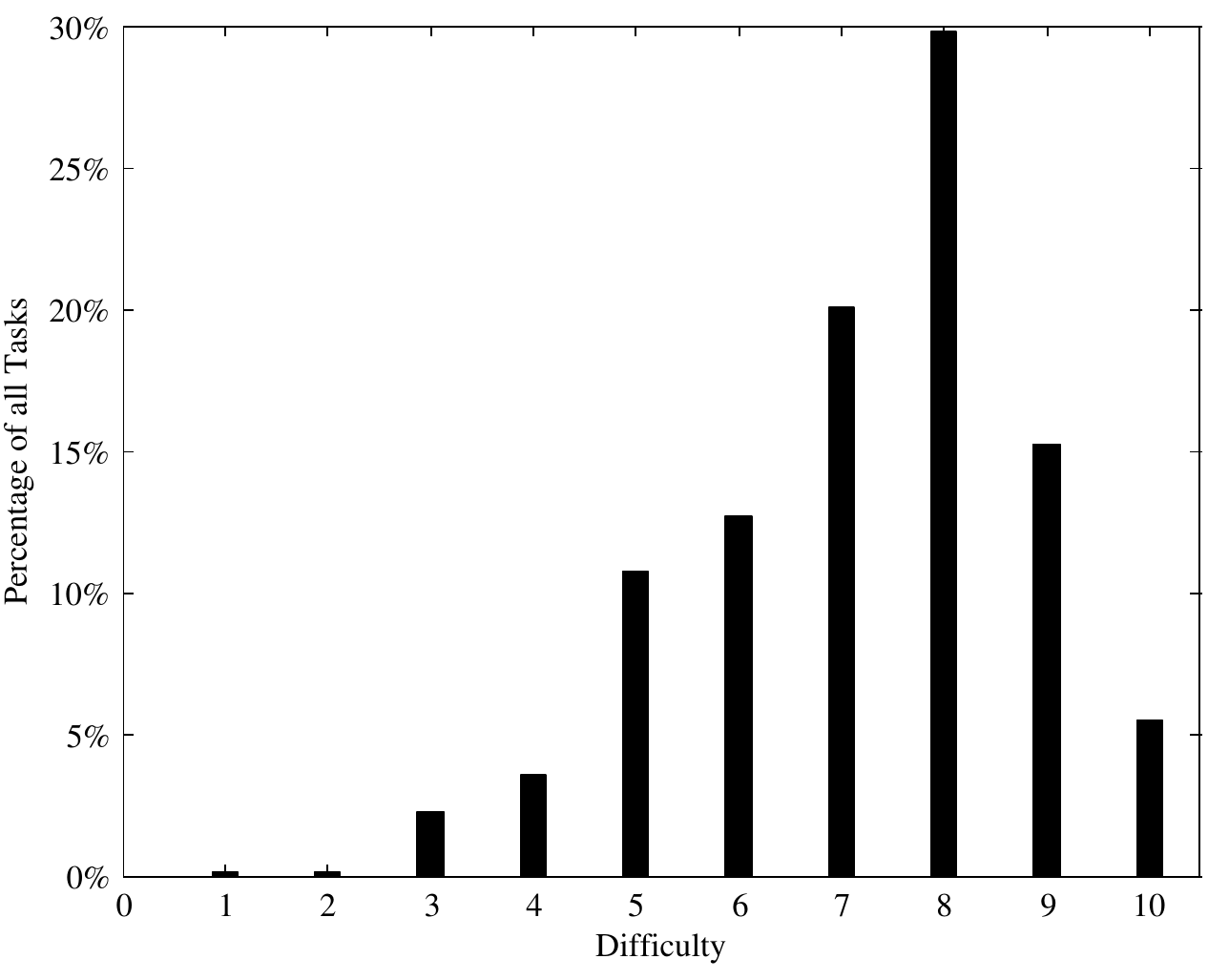}\label{fig:DiffDistribution}}
    \subfigure[The distribution of task deadlines]{
    \includegraphics[trim = 40mm 85mm 45mm 90mm, clip, width = 2.2in]{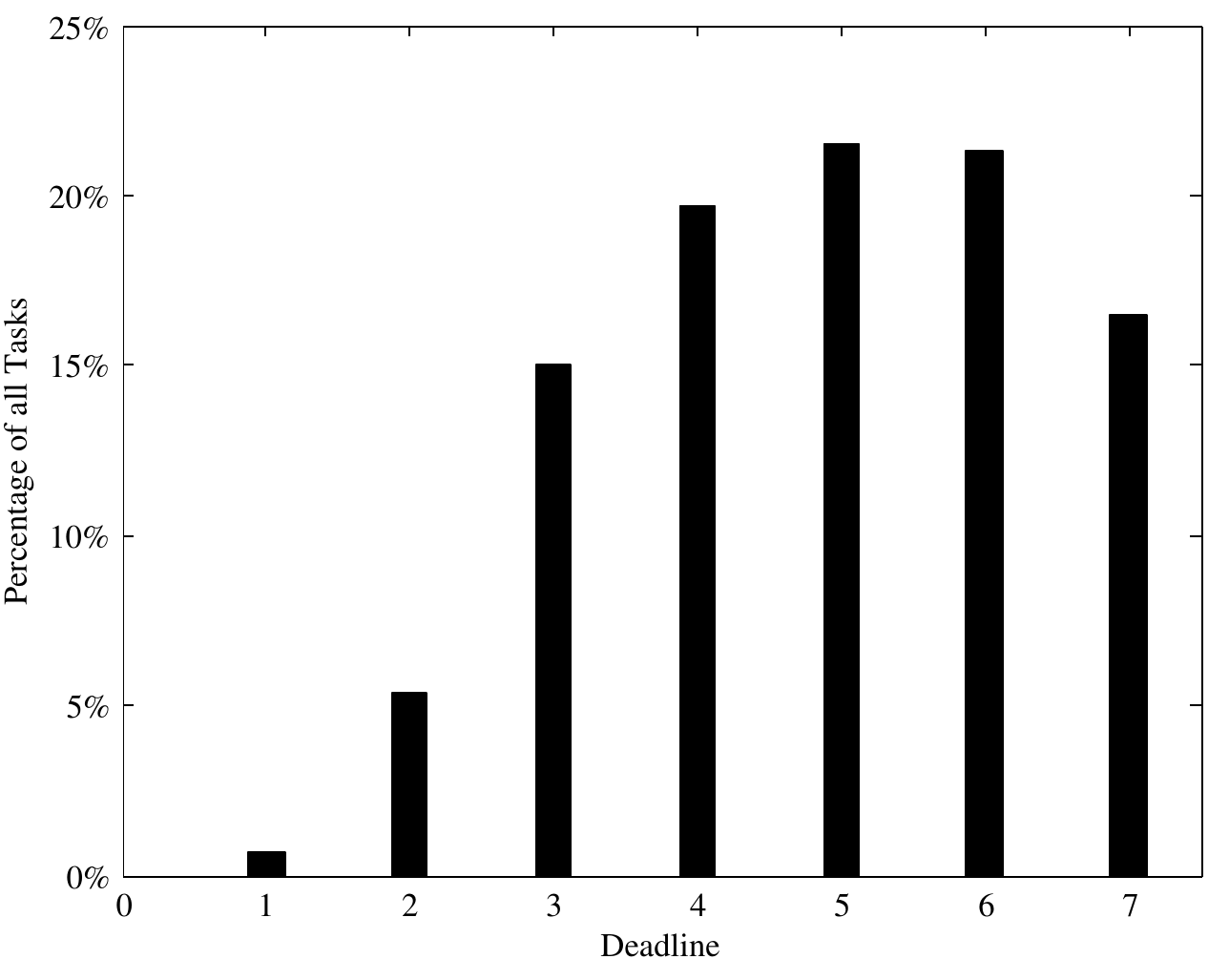}\label{fig:DLDistribution}}
    \caption{Subject Characteristics.} 
\end{figure*}

In this paper, we adopt the exploratory data analysis (EDA) approach \cite{Tukey:1977} to analyze the data collected. EDA is an approach for analyzing data sets to summarize their main characteristics, often with visual methods. It is primarily for understanding what can be learnt from the data beyond the formal modeling or hypothesis testing task. We use the following metrics to facilitate our analysis:
\begin{enumerate}
  \item \textit{Technical Productivity} ($\mu_{i}$): it refers to the average amount of workload a student $i$ can complete during a sprint. In this study, we use the task difficulty value as an indicator of the workload of a task as the task difficulty values reported by students denote both the technical challenge and the amount of effort required to complete the task.
  \item \textit{Competence} ($Comp_{i}$): it refers to the probability a student $i$ can complete a task assigned to him/her with satisfactory quality before the stipulated deadline. In this paper, the outcome of a task needs to achieve an average quality rating higher than 5 out of 10 in order to be considered as having satisfactory quality. This metric is similar to a student's reputation. Thus, we adopt a reputation computation model - the Beta Reputation model \cite{Josang-et-al:2007} - which is widely used in the fields of online services, artificial intelligence and network communications \cite{Pan-et-al:2009,Yu-et-al:2010,Yu-et-al:2011,Yu-et-al:2013}. It is calculated as follows:
      \begin{equation}
        Comp_{i}=\frac{\alpha_{i}+1}{(\alpha_{i}+1)+(\beta_{i}+1)}\in (0,1) \label{eq:1}
    \end{equation}
    where $\alpha_{i}$ and $\beta_{i}$ are calculated as:
    \begin{equation}
    \alpha_{i}= \sum_{\tau\in \phi(i)}1_{[T_{\tau}^{act}-T_{\tau}^{est}\leq 0 \mbox{ and } Qual_{\tau}>5]}D_{\tau} \label{eq:2}
    \end{equation}
    \begin{equation}
    \beta_{i}= \sum_{\tau\in \phi(i)}1_{[T_{\tau}^{act}-T_{\tau}^{est}>0 \mbox{ or } Qual_{\tau}\leq 5]}D_{\tau}. \label{eq:3}
    \end{equation}
    The function $1_{[\mbox{condition}]}$ in Eq. (\ref{eq:2}) and Eq. (\ref{eq:3}) equals to 1 if ``condition" is true. Otherwise, $1_{[\mbox{condition}]}$ equals to 0. $\phi(i)$ denotes the set of tasks $i$ has previously worked on until the current point in time. The ``+1" terms in the numerator and denominator of Eq. (\ref{eq:1}) are \textit{Laplace smoothing} terms \cite{Wang-Singh:2007} which ensure that if $i$ has no previous track record, $Comp_{i}$ evaluates to 0.5 indicating maximum uncertainty about $i$'s performance.
  \item \textit{Workload} ($w_{i}(t)$): it refers to the actual amount of workload assigned to a student $i$ during sprint $t$.
  \item \textit{Final Score} ($s_{i}^{f}$): it refers to the final score a student $i$ achieves for this course. It ranges from 0 to 100 marks.
  \item \textit{Team Score} ($s_{j}$): it refers to the score given to a team $j$ by the course instructor based on the assessment of the software produced by the team at the end of the course work project. It ranges from 0 to 30 marks.
  \item \textit{Collaborators/Task} ($c_{\tau}$): it refers to the number of students working on a same task $\tau$.
  \item \textit{Team Morale (Begin)} ($M_{j}^{begin}(t)$): it refers to the average of the mood values reported by members of team $j$ during the sprint planning meeting of sprint $t$.
  \item \textit{Team Morale (End)} ($M_{j}^{end}(t)$): it refers to the average of the mood values reported by members of team $j$ during the sprint review/retrospective meeting of sprint $t$.
\end{enumerate}

\subsection{Subject Characteristics}
Before the commencement of their course work projects, students involved in our study did not have any experience practising ASD methodologies. Nevertheless, they have received standard training in software engineering concepts in their first year of undergraduate study. The distribution of the students' competence and technical productivity is shown in Figure \ref{fig:StudentsCompCap}. The Pearson Correlation Coefficient (PCC) \cite{Lin:1989} between students' competence and their respective technical productivity values is ($r=0.7443$, $p<0.01$), indicating a statistically significant positive correlation. This means students involved in our study who often complete the tasks assigned to them on time also tend to do so with high quality, and vice versa. As illustrated in Figure \ref{fig:CompDistribution}, the percentages of student population showing various competence levels roughly follow a bell curve centred around 0.4 to 0.5. The distribution of the students' final scores, which include their examination test scores together with their course work project scores, is shown in Figure \ref{fig:FinalScores}. The final scores range from 74 to 100 marks with the majority of students scoring between 80 and 90 marks. As can be observed in Figure \ref{fig:TeamsCompCap}, it turns out that the distribution of the teams' competence and technical productivity is similar to the distribution of individual students' competence and technical productivity in Figure \ref{fig:StudentsCompCap}. As the students decide among themselves on which team they wish to join, we have no control of team characteristics in this study.

Figures \ref{fig:DiffDistribution} and \ref{fig:DLDistribution} shows the distributions of the difficulty values and the deadlines of the 893 tasks proposed by students during this study. The task deadline values range from 1 to 7 days. The task difficulty and deadline roughly follow bell curves centred around 8 and 5, respectively. The PCC between task difficulty and deadline is ($r=0.4086$, $p<0.01$), indicating a statistically significant positive correlation.

\section{Results and Analysis} \label{st:results}

In this section, we present the results from preliminary analysis of the data collected from this study. We focus on three aspects of the Scrum methodology which are important to understanding the agile process and have not been well studied by previous research. They are: 1) task allocation decision-making, 2) collaboration, and 3) team morale.

\subsection{Task Allocation Decision-making}
In \cite{Lin-et-al:2014}, the ratio between participants' competence and the normalized difficulty values of the tasks assigned to them has been shown to positively correlate to the timeliness of task completion. In the data collected, no task was rated by participants as having a difficulty value of 0. Thus, the ratio $\frac{Comp_{i}}{D_{\tau}}\in (0,10)$. If $\frac{Comp_{i}}{D_{\tau}}>1$, it means that a student $i$ is assigned a task $\tau$ with a normalized difficulty value lower than $i$'s competence value. If $\frac{Comp_{i}}{D_{\tau}}\leq 1$, $i$ is assigned a task $\tau$ with a normalized difficulty value higher than or equal to $i$'s competence value. In this study, we investigate whether students take the $\frac{Comp_{i}}{D_{\tau}}$ ratio into account when allocating tasks among themselves.

\begin{figure}
    \centering
    \includegraphics[trim = 40mm 85mm 45mm 90mm, clip, width = 3.3in]{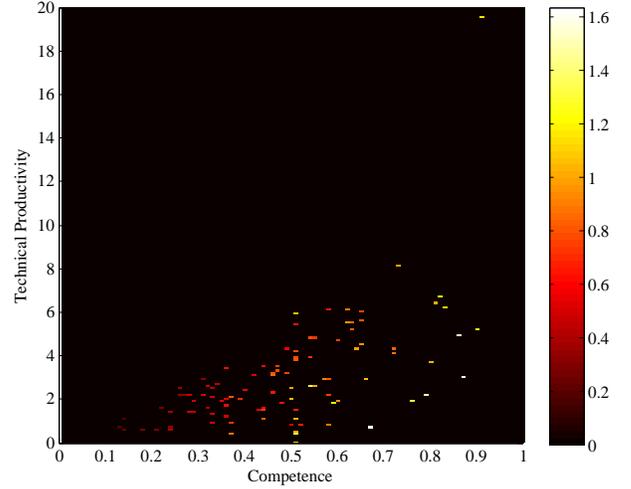}

    \caption{The $\frac{Comp_{i}}{D_{\tau}}$ distribution among students in the study.} \label{fig:CompDiffDistribution}
\end{figure}

Figure \ref{fig:CompDiffDistribution} illustrates the average $\frac{Comp_{i}}{D_{\tau}}$ ratios for all students against their competence and technical productivity. The colour scale corresponds to different $\frac{Comp_{i}}{D_{\tau}}$ ratios. It can be observed that students showing higher competence values tend to be assigned tasks with normalized difficulty values lower than their competence values. The PCC between students' competence values and their $\frac{Comp_{i}}{D_{\tau}}$ ratios in this study is ($r=0.6567$, $p<0.01$), indicating a statistically significant positive correlation. The PCC between students' technical productivity values and their $\frac{Comp_{i}}{D_{\tau}}$ ratios in this study is ($r=0.2992$, $p<0.01$), indicating a statistically significant, albeit weak, positive correlation. Thus, from this study, it appears students indeed attempt to allocate tasks within the assignees' competence when following the Scrum methodology.

\begin{figure}
    \centering
    \includegraphics[trim = 40mm 85mm 45mm 90mm, clip, width = 3.3in]{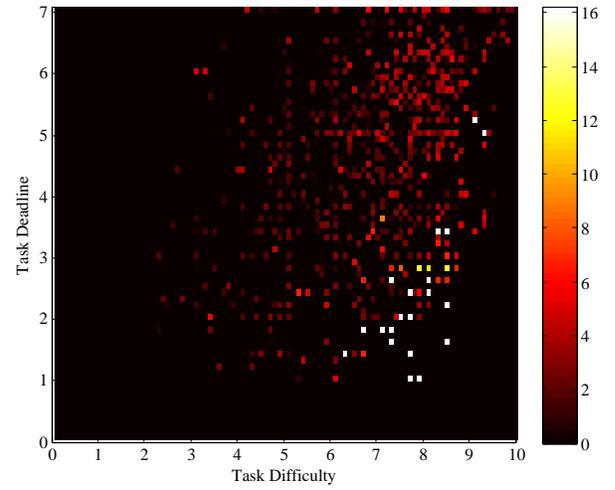}

    \caption{Task allocation v.s. students' technical productivity.} \label{fig:CapacityDiffDL}
\end{figure}

Apart from the $\frac{Comp_{i}}{D_{\tau}}$ ratio, students' technical productivity also plays an important part in task allocation decision-making. Figure \ref{fig:CapacityDiffDL} shows the assignees' technical productivity each task is allocated to. The colour scale corresponds to different technical productivity values. It can be observed that students with high technical productivity are generally allocated more difficult tasks. Tasks with high difficulty values and short deadlines tend to be allocated to students with high technical productivity. The PCC between the $\frac{D_{\tau}}{T_{\tau}^{est}}$ ratio of the tasks and the technical productivity values of the students assigned the tasks is ($r=0.4397$, $p<0.01$), indicating a statistically significant positive correlation.

\subsection{Collaboration}
In this part of the study, we investigate two research questions: 1) what is the relationship between collaboration and team characteristics? and 2) what is the relationship between collaboration and team performance?

\begin{figure}
    \centering
    \includegraphics[trim = 40mm 85mm 45mm 90mm, clip, width = 3.3in]{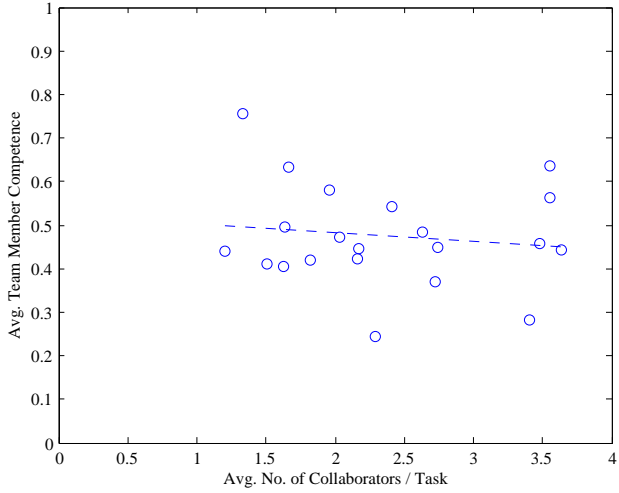}

    \caption{Collaboration v.s. team members' competence.} \label{fig:CollaborationComp}
\end{figure}

With regard to the first research question, we look into the relationship between each team's capabilities and their collaboration behaviours. Figure \ref{fig:CollaborationComp} shows the relationship between the average team competence (which is calculated by averaging team members' competence values) and the average number of collaborators per task in each team. The PCC between the average team competence and the average number of collaborators per task is ($r=-0.1376$, $p=0.5537$), which is not statistically significant. This result favours the null hypothesis that there is not correlation between these two factors.

\begin{figure}
    \centering
    \includegraphics[trim = 40mm 85mm 45mm 90mm, clip, width = 3.3in]{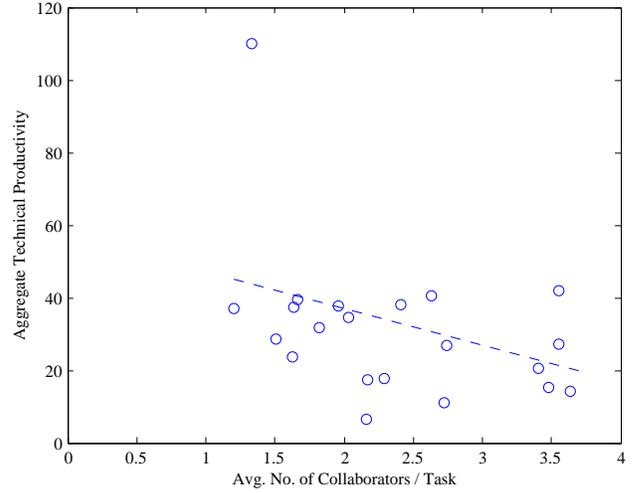}

    \caption{Collaboration v.s. team members' task processing capacity.} \label{fig:CollaborationCap}
\end{figure}

Figure \ref{fig:CollaborationCap} shows the relationship between the aggregate team technical productivity (which is calculated by summing team members' technical productivity values) and the average number of collaborators per task in each team. The PCC between the aggregate team technical productivity and the average number of collaborators per task is ($r=-0.4064$, $p<0.1$), indicating a statistically significant negative correlation. Thus, it appears students in teams with low aggregate team technical productivity values tend to engage in collaborations more often, which is a rational strategy from the students' perspective.

\begin{figure}
    \centering
    \includegraphics[trim = 40mm 85mm 45mm 90mm, clip, width = 3.3in]{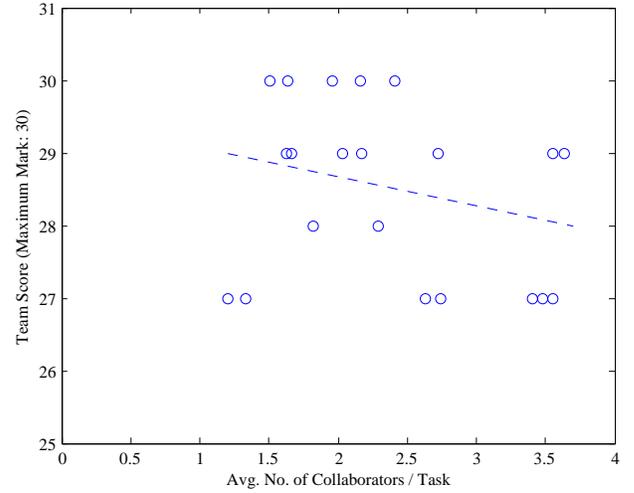}

    \caption{Collaboration v.s. team score.} \label{fig:CollaborationScore}
\end{figure}

In this study, we use the team score to measure the performance of a team. Figure \ref{fig:CollaborationScore} illustrates the relationship between collaboration and team score. The PCC between the team scores and the average number of collaborators per task is ($r=-0.3721$, $p<0.1$), indicating a statistically significant negative correlation. This result contradicts the popular preconception that collaboration improves team performance. Therefore, we conducted an interview with the course instructor to obtain his opinions on the possible reasons for such a negative correlation. The course instructor suggested that due to the lack of team-based project development experience among the students as well as possible differences in terms of competence and technical productivity among team members, collaboration might not have occurred in an effective manner. Another reason for this result can be the incompatibility between collaboration and modularity in software development. In another study from \cite{Kilamo-et-al:2014}, it has been shown that groups with small file-wise collaborative editing ratio tend to score higher grades for the software developed.

\subsection{Team Morale}
\begin{figure}
    \centering
    \includegraphics[trim = 40mm 85mm 45mm 90mm, clip, width = 3.3in]{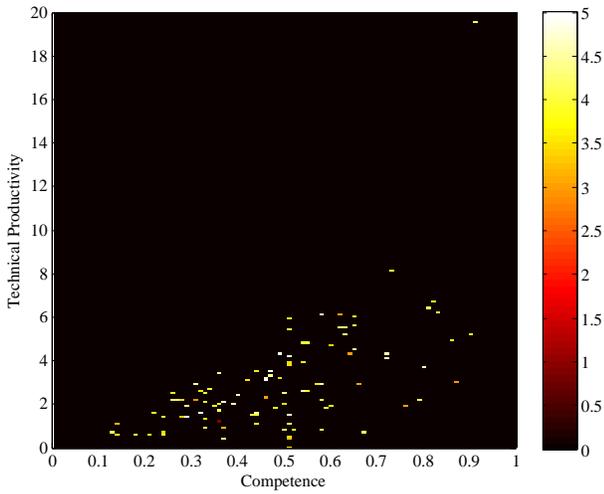}

    \caption{Students' average morale before a Sprint.} \label{fig:Morale(Start)}
\end{figure}

Figure \ref{fig:Morale(Start)} shows the distribution of students' average self-reported mood values during the sprint planning meeting at the start of each sprint. The colour scale represents the average self-reported mood values. The average mood value is 3.86 out of 5. The PCC between students' mood during the sprint planning meetings and their competence values is ($r=-0.0025$, $p=0.9394$), indicating no statistically significant correlation. The PCC between students' mood during the sprint planning meetings and their technical productivity values is ($r=0.1505$, $p<0.01$), indicating a statistically significant albeit weak positive correlation.

\begin{figure}
    \centering
    \includegraphics[trim = 40mm 85mm 45mm 90mm, clip, width = 3.3in]{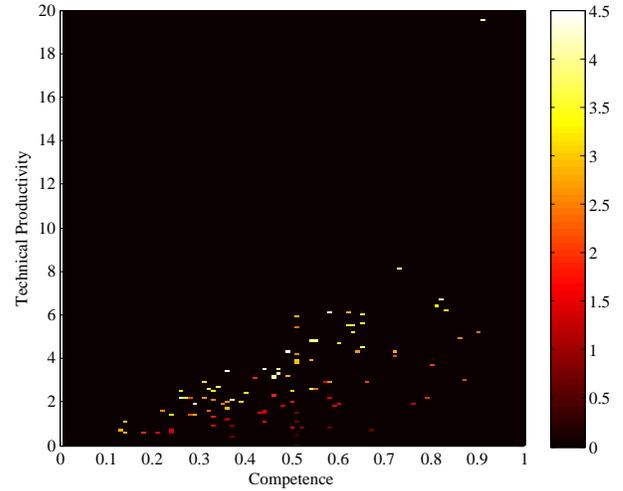}

    \caption{Students' average morale after a Sprint.} \label{fig:Morale(End)}
\end{figure}

Figure \ref{fig:Morale(End)} shows the distribution of students' average self-reported mood values during the sprint review/retrospective meeting at the end of each sprint. The colour scale represents the average self-reported mood values. The average mood value is 3.80 out of 5 which is slightly lower than at the beginning of the sprint. The PCC between students' mood during the sprint review/retrospective meetings and their competence values is ($r=-0.0148$, $p=0.5946$), indicating no statistically significant correlation. The PCC between students' mood during the sprint review/retrospective meetings and their technical productivity values is ($r=0.4207$, $p<0.01$), indicating a statistically significant positive correlation. Therefore, based on these analysis, team members with high technical productivity tend to have high morale, especially at the end of a sprint after completing the tasks allocated to them.

\begin{figure}
    \centering
    \includegraphics[trim = 40mm 85mm 45mm 90mm, clip, width = 3.3in]{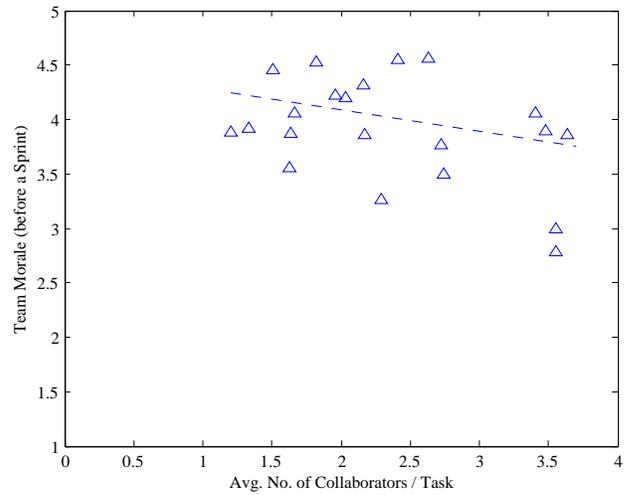}

    \caption{Collaboration v.s. average team morale before a Sprint.} \label{fig:CollaborationMorale(Start)}
\end{figure}

In addition to investigating the relationship between students' capabilities and their morale, we also investigate the relationship between collaboration and team morale. The morale value of a team is the average of the mood values reported by its members over the 12 week period. Figure \ref{fig:CollaborationMorale(Start)} shows the relationship between the team morale values during the sprint planning meetings and the average number of collaborators per task in each team. The PCC between team morale during the sprint planning meetings and the average number of collaborators per task is ($r=-0.4135$, $p<0.1$), indicating a statistically significant negative correlation.

\begin{figure}
    \centering
    \includegraphics[trim = 40mm 85mm 45mm 90mm, clip, width = 3.3in]{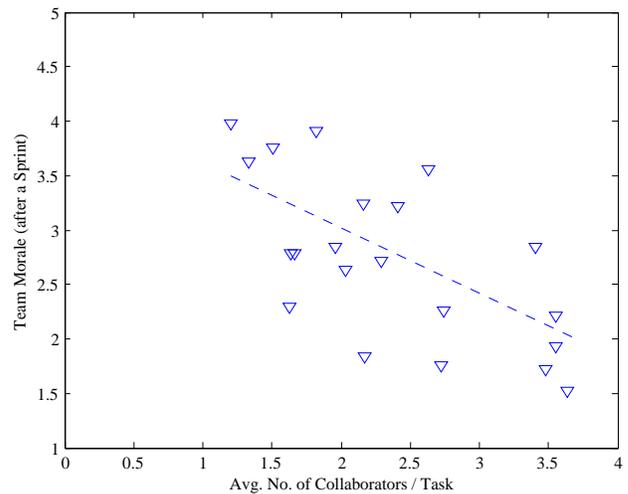}

    \caption{Collaboration v.s. average team morale after a Sprint.} \label{fig:CollaborationMorale(End)}
\end{figure}

Figure \ref{fig:CollaborationMorale(End)} shows the relationship between the team morale values during the sprint review/retrospective meetings and the average number of collaborators per task in each team. The PCC between team morale during the sprint review/retrospective meetings and the average number of collaborators per task is ($r=-0.6632$, $p<0.01$), indicating a statistically significant negative correlation. Therefore, the results suggest that collaboration by team members who are inexperienced in software development in Scrum teams negatively affects team morale.

In summary, the key findings about the Scrum-based ASD process practised by novice teams from this study are: 1) task allocation in agile teams positively correlate to students' technical productivity; 2) collaboration is negatively correlated with team technical productivity, team morale, and team score; 3) team morale is positively correlated to their technical productivity.

\section{Study Limitations} \label{st:limitations}
This section discusses the study limitations based on three categories of threats to validity described in \cite{Wohlin-et-al:2000}. For each category, we list all possible threats, measures taken to reduce the risks, and suggestions for improvements in future studies.

\subsection{Internal Validity}
Internal validity threats that may have affected our study are the lack of control of the following variables: 1) the students' competence (other than all being in the same semester of the course); and 2) how the teams are formed (students decide on their own which teams they wish to join, the instructor only controls the sizes of the teams). With respect to Threat 1, the repeated Scrum activity data logging over a 12 week period of time decreases the probability of this threat affecting our outcomes to some extent as students are provided with many opportunities to demonstrate their competence on different tasks. Threat 2 has affected the study. In regard to it, we believe the sample size, which was not small (125 students performing close to 900 tasks over 12 weeks), reduces the extent of the effect of this threat (i.e., different types of pairings with respect to student competence and task difficulty have occurred). In future studies, we suggest that pre-testing should be organized to assess the competence of the students. The results can be used to guide instructors in organizing the students into teams.

\subsection{External Validity}
A factor that might reduce the external validity of our study is the use of students as subjects. Nevertheless, according to \cite{Basili-et-al:1999,Kitchenham-et-al:2002}, students can play an important role in experimentation in the field of software engineering. Some attempts to replicate the same studies with both student and professional subjects even produced similar results \cite{Canfora-et-al:2007}. However, to be conservative, we refrain from generalizing our results to professionals practising Agile programming. In future studies, we plan to replicate the experiment with professionals.

Another threat to the external validity of our study is the representativeness of the tasks proposed for the students to work on. As tasks are proposed by each team during their Sprint planning meetings based on the objectives of their respective projects, we do not have control over the types, priorities, difficulties, and expected deadlines of the tasks involved in the study. However, such a situation is similar to what happens in real world Agile programming, and, thus, its impact on the validity of the study should not be over-emphasized. In any case, the large scale and long period of time of this study is one way to reduce the effect of this threat. In the future, we plan to replicate our experiments with more well defined tasks of various complexity (possibly from open source software projects).

\subsection{Construct Validity}
A characteristic of our study that might affect its construct validity is that students had limited previous experience with the Scrum ASD approach during the course work. In addition, in most cases, students in the same team had not worked with their team mates before. Therefore, similar to \cite{Lemos-et-al:2012}, our results might be conservative with respect to the effects of collaboration. In subsequent studies, we will consider involving programmers who have more experience with this development approach and who have worked together before.

\section{Conclusions and Future Work} \label{st:conclusions}
Existing empirical studies mostly focus on XP-based ASD methodologies. The recent significant rise in popularity of Scrum-based ASD methodologies demands related empirical studies to produce insights that can guide the design of future training programmes. The study reported in this paper, which involves 125 students over 12 weeks, is a timely response to this demand from the research community. Different from traditional survey/interview-based empirical studies, our study is based on participants' ASD activity trajectory data collected unobtrusively during normal ASD processes through our HASE APM platform. This type of data objectively reflects users' ASD activities and performance at fine granularities. With the help of this form of data, we report key findings in this paper that reveal important insights into the Scrum ASD processes when practised by novice student teams. These results offer new insights into the aspects of agile team collaboration and team morale which have not yet been well studied by existing research.

With this study, we see the start of a series of research on agile software development with ASD activity trajectory data. In future research, we will explore various modeling approaches (e.g., fuzzy cognitive approaches \cite{Miao-et-al:2002,Song-et-al:2009}, evolutionary methods \cite{Li-et-al:2009}, and inference models \cite{Miao-et-al:2001}) to design personalized inference models to convert the ASD team members' behavior trajectory data into predictive analytics models for task allocation decision support. Such models may be combined with game-based elements to design training environments \cite{Michael-et-al:2010} to improve ASD team members' task allocation skills. We also plan to conduct surveys/interviews to understand more in-depth how students in each Scrum team collaborate. We will continue using the HASE platform to collect agile programming activity data over subsequent semesters and expand our data collection effort to include more universities so as to investigate the possible effects of socio-cultural factors. The resulting datasets will be published in the future to support the discovery of new insights by researchers in the field. We will also propose machine learning methods to analyze the datasets to look for latent factors that might have been missed by EDA.


\section*{Acknowledgment}
This research is supported by the  National Research Foundation, Prime Minister's Office, Singapore under its IDM Futures Funding Initiative and administered by the Interactive and Digital Media Programme Office.



%

\bibliographystyle{IEEEtran}
\bibliography{Reference}

\end{document}